\documentclass[osajnl,twocolumn,showpacs,superscriptaddress,10pt]{revtex4-1} 
\usepackage{amsmath,amssymb,graphicx}
\newcommand{\mic}{~\mu \rm m}
\DeclareMathOperator{\sech}{sech}
\begin{document}

\title{
Mid-IR femtosecond frequency conversion by soliton-probe collision in phase-mismatched quadratic nonlinear crystals}

\author{Xing Liu}
\affiliation{DTU Fotonik, Dept. of Photonics Engineering, Technical University of Denmark, DK-2800 Kgs. Lyngby, Denmark}

\author{Binbin Zhou}
\affiliation{DTU Fotonik, Dept. of Photonics Engineering, Technical University of Denmark, DK-2800 Kgs. Lyngby, Denmark}

\author{Hairun Guo}
\affiliation{DTU Fotonik, Dept. of Photonics Engineering, Technical University of Denmark, DK-2800 Kgs. Lyngby, Denmark}
\affiliation{Present address: Ecole Polytechnique Federale de Lausanne, CH-1015 Lausanne, Switzerland}

\author{Morten Bache}
\affiliation{DTU Fotonik, Dept. of Photonics Engineering, Technical University of Denmark, DK-2800 Kgs. Lyngby, Denmark}
\affiliation{Corresponding author: moba@fotonik.dtu.dk}

\begin{abstract} We show numerically that ultrashort self-defocusing temporal solitons colliding with a weak pulsed probe in the near-IR can convert the probe to the mid-IR. A near-perfect conversion efficiency is possible for a high effective soliton order. The near-IR self-defocusing soliton can form in a quadratic nonlinear crystal (beta-barium borate) in the normal dispersion regime due to cascaded (phase-mismatched) second-harmonic generation, and the mid-IR converted wave is formed in the anomalous dispersion regime between $\lambda=2.2-2.4\mic$ as a resonant dispersive wave. This process relies on non-degenerate four-wave mixing mediated by an effective negative cross-phase modulation term caused by cascaded soliton-probe sum-frequency generation.
\end{abstract}


\maketitle 

\noindent
The optical soliton is remarkably robust as it can both retain its shape despite dispersive or dissipative effects and survive wave collisions. Yet it is quite susceptible to perturbations as it can shed phase-matched resonant radiation to a so-called soliton-induced optical Cherenkov wave (a.k.a. dispersive wave) when perturbed by higher-order dispersion \cite{Wai:1986,beaud:1987,Wise:1988,Gouveia-Neto:1988,Skryabin:2010}. On the other hand, the soliton can also act as a potential barrier when colliding with a linear (i.e. dispersive) wave, creating the analogy to an optical equivalent of an "event horizon" \cite{Philbin:2008,demircan:2011}. Such a collision can be well understood by generalizing the Cherenkov phase-matching condition -- degenerate four-wave mixing (FWM) -- to non-degenerate FWM where a soliton interacts with two linear dispersive waves
\cite{Yulin:2004,skryabin:2005,Efimov2005,Genty2012-APC}. This interaction is mediated by cross-phase modulation (XPM): the collision between the soliton ($\omega_s$) and a linear "probe" wave ($\omega_p$) can become resonantly phase-matched to a new frequency (the "resonant" wave $\omega_r$) according to the FWM phase-matching condition. This frequency-converts the probe to the resonant wave, which -- when completely depleting the probe --  gives rise to the peculiar appearance of the probe reflecting on the soliton: when the probe group velocity is higher than the soliton, the resonant wave group velocity will be lower than the soliton, and therefore travel away from the soliton after formation. This frequency-conversion process therefore in time domain leads to an apparent reflection of the probe on the soliton. 


Soliton-probe collisions have been studied in fibers, which have a positive Kerr nonlinearity. However, through cascaded (strongly phase-mismatched) quadratic nonlinear interactions an effective negative Kerr-like nonlinearity may be generated (in bulk this corresponds to a self-defocusing effect) \cite{desalvo:1992}. As a consequence soliton formation requires normal dispersion \cite{ashihara:2002,zhou:2012}, and Cherenkov phase-matching naturally occurs towards the red side of the soliton spectrum \cite{bache:2010e}. This allows for efficient near- to mid-IR conversion \cite{bache:2011a,Zhou:2015}.

We here study the collision of near-IR probe waves and self-defocusing solitons to generate long-wavelength resonant waves. The self-defocusing soliton may reflect the probe wave when the probe XPM term is negative, and this is possible if the sum-frequency generation (SFG) between soliton and probe is detuned sufficiently away from its phase-matching point to induce a negative cascaded XPM term. We find that complete conversion from the probe to the resonant wave requires interaction with a higher-order soliton, and demonstrate a wide tunability of the wavelength of the mid-IR resonant wave by varying the probe center wavelength.

\begin{figure}[tb]
\begin{centering}
\includegraphics[width=0.988\linewidth]{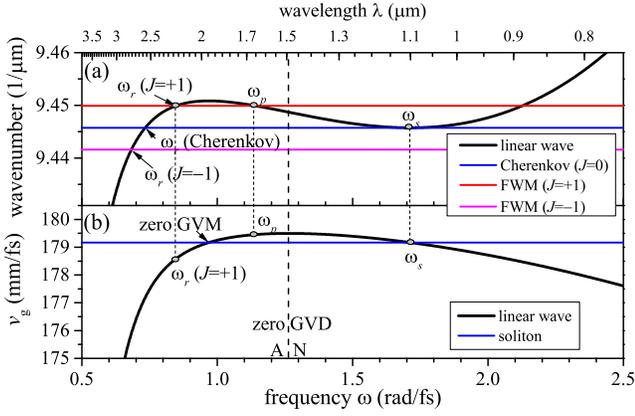}
\end{centering}
  \caption{(a) The dispersion relations for BBO in frequency domain, reported in the soliton group-velocity frame. Black line: left-hand side of Eq. (\ref{eq:PM}), colored lines: right-hand side of Eq. (\ref{eq:PM}), taking a soliton at $\lambda_s=1.1\mic$ and a probe at $\lambda_p=1.65\mic$. "A" and "N" denote regimes with anomalous and normal dispersion, respectively. 
  (b) The group velocities of a linear wave (black) and the soliton at $\lambda_s=1.1\mic$ (blue).
  }
  \label{fig:PM}
\end{figure}

The general FWM phase-matching condition 
is \cite{skryabin:2005}
\begin{align}\label{eq:PM}
  k_{\rm lin}(\omega_r)=k_{\rm sol}(\omega_r)+J[k_{\rm lin}(\omega_p)-k_{\rm sol}(\omega_p)]
\end{align}
where $k_{\rm lin}(\omega)$ describes the dispersion relation of the linear wave (in bulk media simply determined, e.g., by the Sellmeier equation). $k_{\rm sol}=k_{\rm lin}(\omega_{s})+(\omega-\omega_{s})/v_{g,\rm sol}+q_{\rm sol}$ is the soliton dispersion relation; its non-dispersive nature is reflected in the fact it is simply a wave packet with a constant group velocity $v_{g,\rm sol}$. Its accumulated nonlinear phase $q_{\rm sol}$ will cancel out for the $J=+1$ case that we will focus on here. The parameter $J$ switches between the degenerate case ($J=0$, Cherenkov radiation) and the non-degenerate case ($J=\pm 1$, where the presence of the probe at frequency $\omega_p$ invokes the FWM resonant phase-matching condition). In a BBO quadratic nonlinear crystal ($\beta$-barium borate, BaB$_2$O$_4$) the resonant waves are phase-matched in the mid-IR beyond $\lambda=2.0\mic$, as seen from the dispersion relations in Fig. \ref{fig:PM} for the main case considered here, namely a $1.65\mic$ probe colliding with a $1.1\mic$ soliton.

The BBO crystal is assumed cut for type-I second-harmonic generation (SHG), where two $o$-polarized photons at the fundamental wave (FW) frequency $\omega_1$ generate a second-harmonic (SH) $e$-polarized photon at the frequency $\omega_2=2\omega_1$. The numerics use the nonlinear wave equations in frequency domain \cite{Guo:2013} model; the values of the $\chi^{(2)}$ and $\chi^{(3)}$ tensor components were chosen from \cite{Bache:2013}, and the Raman effect is neglected as it is usually considered weak in BBO. Here we pump in the $o$-wave and through phase-mismatched SHG to the $e$-wave a nonlinear phase shift accumulates on the pump pulse, which we exploit to excite a self-defocusing soliton. Importantly, since we also pump with a weak $o$-polarized probe wave, the numerical model also includes any possible $\chi^{(2)}$ interaction, such as sum- and difference frequency generation (SFG and DFG), and both inter- and intra-polarization (i.e. type 0, I and II) interactions. For simplicity the BBO mid-IR material loss is neglected. 

The collision is modelled by launching two co-propagating $o$-polarized fields 
$E_{o,\rm in} =E_{1, \rm in}\cos(\omega_1 t)\sech(t/T_1)+E_{p, \rm in} \cos(\omega_p t)\sech[(t-\tau)/T_p]$, where ${\tau}$ is the delay time between them and we will use identical input pulse durations $T_1=T_p$.
A self-defocusing soliton can form at $\omega_s=\omega_1$ if the following criteria are fulfilled (see \cite{Liu:2015-BBO} for more details): (a) The effective Kerr self-phase modulation (SPM) nonlinearity $n_{2,\rm eff}^{\rm SPM}(\omega_1)=n^{\rm SHG}_{2, \rm casc}(\omega_1)+n_{2,\rm Kerr}$ must be negative, and this is controlled by $n_{2, \rm casc}^{\rm SHG}(\omega_1)\propto -d_{\rm eff}^2/\Delta k_{\omega_1}^{\rm SHG}$ by making the SHG phase mismatch $\Delta k_{\omega_1}^{\rm SHG}=k_e(2\omega_1,\theta)-2k_o(\omega_1)$ small enough; (b) the group-velocity dispersion (GVD) must be normal [$k_{o}^{(2)}(\omega_1)>0$], which in BBO means $\lambda_1<1.488\mic$; (c) the effective soliton order \cite{bache:2007} $N_{\rm eff}\geq 1$, where ${ N_{\rm eff}^2=L_D\omega_1 I_{\rm in} |n_{2, \rm eff}^{\rm SPM}(\omega_1)|/c }$, $I_{1,\rm in}=\varepsilon_0 n_1(\omega_1) c|E_{1,\rm in}|^2/2$ 
and ${ L_D=T_1^2/ k_{o}^{(2)}(\omega_1)}$.

The soliton will reflect/scatter the probe if the soliton-induced XPM potential is a barrier. As the probe is launched in the anomalous dispersion regime, this requires a negative probe XPM effective nonlinear index. This has the contributions \cite{Liu:2015-BBO}: $n_{2, \rm eff}^{\rm XPM}=n_{2, \rm Kerr}^{\rm XPM}+n_{2, \rm casc}^{\rm SFG}(\omega_s+\omega_p,\theta)+n_{2, \rm casc}^{\rm DFG}(\omega_s-\omega_p,\theta)$, viz. the material Kerr XPM (which in BBO is identical to $n_{2, \rm Kerr}^{\rm SPM}$, as both waves are $o$-polarized), as well as type I cascaded soliton-probe SFG and DFG. These are similar in form to $n_{2, \rm casc}^{\rm SHG}$ \cite{Liu:2015-BBO} and in our case we find that $n_{2, \rm casc}^{\rm DFG}(\omega_s-\omega_p,\theta)$ is negligible. Thus, the material Kerr XPM effect and the cascaded SFG effect are the main contributions to the XPM sign and magnitude. Only in certain regimes may $n_{2, \rm eff}^{\rm XPM}<0$ \cite[Fig. 6]{Liu:2015-BBO}, and the potential may even flip sign to become a hole.

\begin{figure}[tb]
\includegraphics[width =\linewidth]{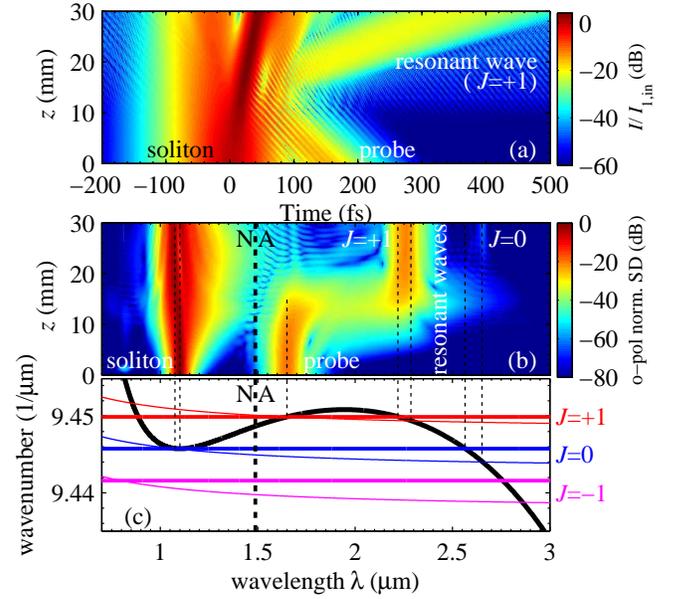}
\caption{Evolution of the $o$-polarized field in (a) time domain (normalized intensity of the electrical field envelope) and (b) wavelength domain. (c) The dispersion curves of Fig. \ref{fig:PM}(a) in wavelength domain for $\lambda_s=1.1\mic$ (thick) and $\lambda_s=1.075\mic$ (thin). Input pulses: 50 fs FWHM@$\lambda_1=1.1\mic$, ${I_{1,\rm in}=200~\rm GW/cm^2}$ (soliton) and 50 fs FWHM@$\lambda_p=1.65\mic$, ${I_{p,\rm in}=5~\rm GW/cm^2}$, $\tau=150$ fs (probe). The BBO crystal had ${\Delta k_{\omega_1}^{\rm SHG}=60~\rm mm^{-1}}$ (${\theta=18.8^\circ}$ and ${\varphi=-90^\circ}$).
}
\label{fig:sim}
\end{figure}

Fig. \ref{fig:sim} shows the results from a typical simulation. 
The BBO crystal angle is suitably chosen to give  nonresonant \cite{bache:2007a} negative SPM and XPM nonlinearities (which occurs between $17.5^\circ <\theta<20.0^\circ$ \cite[Fig. 6]{Liu:2015-BBO}). The soliton input intensity is chosen so $N_{\rm eff}=2.0$, allowing a higher-order self-defocusing soliton to form. 
A weak probe is launched in the anomalous dispersion regime, which from Fig. \ref{fig:PM}(b) implies that its group velocity is larger than the soliton. It is therefore suitably delayed at the input so the interaction occurs over realistic crystal lengths. The time plot in (a) shows the probe colliding with the trailing edge of the strong soliton at around 10 mm. After the collision a reflected wave emerges; this is the resonant wave phase-matched to the soliton through the negative XPM nonlinearity. According to Fig. \ref{fig:PM}(b) the resonant wave will have a lower group velocity than the soliton and this explains why it is traveling away from the soliton trailing edge. In wavelength domain (b) the normalized spectral density (SD, calculated as $S(\lambda)=|\tilde A(\omega=2\pi c/\lambda)|^2\lambda^2/c$) shows that the probe is almost completely converted to the resonant wave, and this occurs between 10 and 20 mm propagation. It is exactly in this propagation range the collision takes place in time domain. There is a good agreement between the predicted phase-matching frequency of the resonant wave, which is evident from the wavelength-domain phase-matching curves plotted in (c).

We note that the soliton blue-shifts slightly during propagation as a consequence of cascading-induced self-steepening. This means that at the final stage $\omega_s>\omega_1$, and this leads to a new set of soliton curves for the phase-matching conditions, indicated as thin lines in (c). As these represent solitons they are in frequency domain still straight curves, but they are now tilted instead of flat as the group velocity is different from that at $\omega_1$; in wavelength domain this means that they are no longer represented as straight curves, but this is simply due to the $\lambda\propto 1/\omega$ relation. These curves explain how the resonant wave is found slightly more red-shifted than the $\omega_s=\omega_1$ case predicted. We also see that the Cherenkov ($J=0$) case can be seen in the spectrum, accurately predicted by the blue-shifted soliton phase-matching condition. This is the degenerate case, where the soliton alone becomes phase-matched to a resonant wave.

\begin{figure}[tb]
\includegraphics[width =\linewidth]{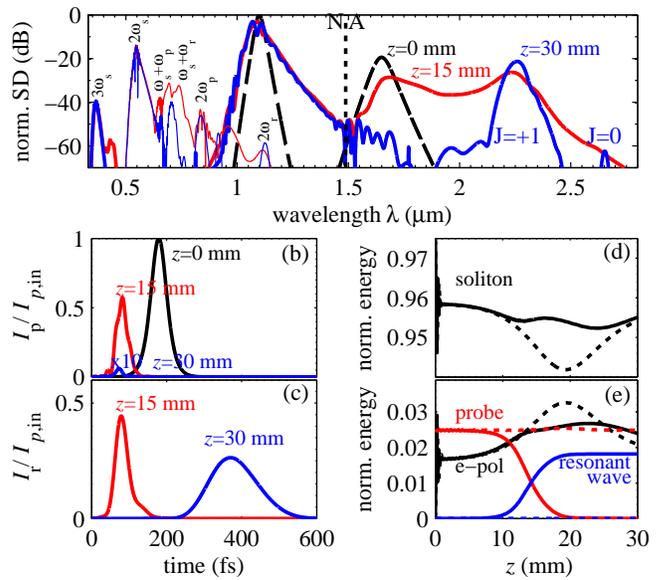}
\caption{Cuts from Fig. \ref{fig:sim}: (a) the normalized $o$- (thick) and $e$-polarized (thin) SD; (b) the band-pass filtered probe and (c) the long-pass filtered resonant wave envelopes vs. time. (d+e) Normalized energy vs. $z$ in the collision (full) and no collision (dashed) cases of (d) the soliton and (e) the entire $e$-polarized spectrum, the filtered probe and resonant waves. 
}
\label{fig:cut}
\end{figure}

Detailed spectral and temporal cuts are shown in Fig. \ref{fig:cut}. (a) shows the SD at input, during ($z=15$ mm) and after collision ($z=30$ mm). Since the soliton order is above unity, the $o$-polarized spectra (thick lines) show that the soliton at collision is considerably extended towards the probe spectrum, and the final spectrum shows that the probe is almost completely depleted leaving only the $J=+1$ and $J=0$ resonant waves. The $o$-polarized third harmonic is also evident, and in the $e$-polarized spectra (thin lines) the various SHG and SFG components are evident as well. The same cuts are shown in time domain focusing in (b) on the probe (using a band-pass filter) and in (c) on the resonant wave (using a long-pass filter). The probe at 15 mm is around half depleted, giving most of its depleted energy to the resonant wave that is located at the same temporal position. After 30 mm the weak probe does not show on a linear scale (in the plot it is amplified 10 times). The resonant wave is now delayed 400 fs and it is reduced in intensity and increased in time due to dispersion. It has a Gaussian profile since it is a linear and not a soliton wave. Finally (d+e) show the energy, normalized to the total input energy, of the soliton, probe and resonant waves. The soliton initially looses around 2\% of its energy through SHG to the $e$-polarized SH [(e) also shows the total $e$-polarized energy], causing the initial ripples at $z<1$ mm. The soliton-probe interaction occurs between $z=5-20$ mm, and the resonant wave builds up in energy. After 20 mm the probe is depleted. The energy ratio (conversion efficiency) of the resonant wave to the probe is around 0.72, close to the limit posed by the photon-to-photon ratio $\omega_r/\omega_p=0.73$ as dictated by the Manley-Rowe relation. The energies from a simulation where the probe never collides with the soliton (dashed lines) show as expected no energy at the resonant wave and the probe remains unaffected. 

\begin{figure}[tb]
\centerline{\includegraphics[width=0.992\linewidth]{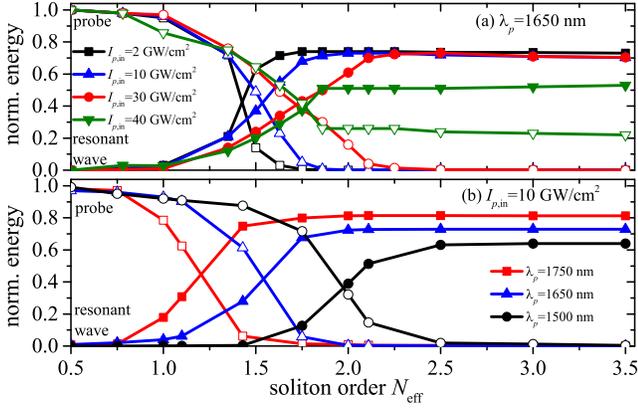}}
\caption{The probe and resonant wave energies (normalized to probe input energy) vs. $N_{\rm eff}$ at $z=40$ mm. (a) Fixing $\lambda_p=1.65\mic$ and (b) fixing $I_{p,\rm in}=10~\rm GW/cm^2$.}
\label{fig:Neff}
\end{figure}%

Fig. \ref{fig:Neff} shows the probe and resonant wave energies vs. $N_{\rm eff}$ (controlled by $I_{1,\rm in}$) for (a) $\lambda_p=1.65\mic$ fixed and four different $I_{p,\rm in}$ values, and (b) $I_{p,\rm in}=10~\rm GW/cm^2$ fixed and three different $\lambda_p$ values. In (a) the complete depletion of the probe happens for intensities up to $30~\rm GW/cm^2$, all ending up at the same plateau, whose level is dictated by the Manley-Rowe relation (as $\omega_p$ is fixed). For $I_{p,\rm in}=40~\rm GW/cm^2$ the probe conversion is incomplete, which is a result of nonlinear spectral broadening occurring  before collision. It is therefore important that the probe does not experience any nonlinear phase shifts before the collision. In (b) a fixed moderate probe intensity was therefore used to ensure complete depletion of the probe. Since different probe frequencies are used the plateau levels vary, in accordance with the Manley-Rowe relation. Clearly probe depletion requires $N_{\rm eff}>1$. Generally the resonant wave growth follows a logistic sigmoid function, whose slope scales as $I_{p,\rm in}^{-0.5}$ and the midpoint $N_{\rm eff,0}$ increases linearly with $I_{p, \rm in}$ and decreases linearly with $\lambda_p$. This latter scaling comes from the fact that as $\lambda_p$ increases it approaches the zero group-velocity mismatch (GVM) wavelength, where the probe and soliton have identical group velocities, see Fig. \ref{fig:PM}(b). The detuning from GVM has traditionally been kept small to get a large reflection, with the dilemma that the resonant wavelength is almost identical to the probe [see again Fig. \ref{fig:PM}(b)] and the collision will only take place through very long interaction distances \cite{Genty2012-APC}. Still a decent conversion should be possible even with large detunings from the zero GVM point (see e.g. \cite{Choudhary2012}), and we intend to investigate this further in another paper.

\begin{figure}[b]
\centerline{\includegraphics[width=\linewidth]{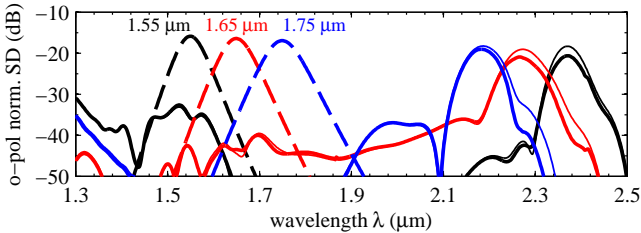}}
\caption{Long-wavelength part of the $o$-polarized SD for various probe wavelengths shown after $z=20$ mm (except $\lambda_p=1.75\mic$ where $z=30$ mm) with (thick) and without (thin) IR losses of BBO. The dashed lines show the input states. The soliton was the same as in Fig. \ref{fig:sim} and $I_p=10~\rm GW/cm^2$.}
\label{fig:probe}
\end{figure}%

The emission wavelength of the mid-IR resonant wave is tunable from $\lambda=2.2-2.4\mic$ through changing the probe wavelength, see Fig. \ref{fig:probe}. Note that the mid-IR wave remains quite significant even when probe-depletion is incomplete and when IR losses are taken into account. Besides, this process can be generalized to mid-IR transparent crystals like LiNbO$_3$ \cite{zhou:2012,Zhou:2015}, where the phase-matching point will also lie further into the mid-IR.

Concluding, in a quadratic nonlinear crystal multiple cascaded nonlinear effects allow firstly the formation of a self-defocusing near-IR soliton (through cascaded SHG generating a self-defocusing Kerr-like SPM term), which when colliding with a weak near-IR probe induces a cascaded SFG nonlinearity generating a self-defocusing Kerr-like XPM term. This allows a resonant mid-IR wave to become phase matched. A complete probe-resonant wave conversion is possible when colliding with a higher-order soliton, a case which has not been considered before in Kerr systems, and for a given soliton wavelength the resonant wavelength is tunable by varying the probe wavelength. Obtaining phase matching further into the mid-IR is possible in other crystals, and interestingly the system allows to change the XPM term sign to study the barrier vs. hole potential effect.


X.L., M.B. and B.Z. acknowledge the support from the Danish Council for Independent Research (11-106702).


\appendix

\section{Theoretical background}

\noindent Let us first stress that the code we use models the $o$- and $e$-polarized electrical fields and include all possible $\chi^{(2)}$ and $\chi^{(3)}$ interactions. However, an analytical understanding of the processes behind the interaction is more convenient in the slowly varying envelope approximation. We will now consider the interaction of a strong soliton envelope $E_s$ at the frequency $\omega_s$ and a weaker (but not weak) probe envelope $E_p$ at the frequency $\omega_p$. Let us focus on the two equations for the soliton and probe and disregard dispersion for the moment. For the soliton
\begin{multline}\label{eq:soliton}
id_z E_{\omega_s}+\frac{\omega_s d_{\rm eff}}{cn_o(\omega_s)}[
E_{\omega_s}^*E_{2\omega_s}e^{i\Delta k_{\omega_s}^{\rm SHG}z}\\
+
E_{\omega_p}E_{\omega_s-\omega_p}e^{-i\Delta k_{\omega_s-\omega_p}^{\rm DFG}z}
\\
+E_{\omega_p}^*E_{\omega_s+\omega_p}e^{i\Delta k_{\omega_s+\omega_p}^{\rm SFG}z}
]
\\+
\frac{3\omega_s}{8n_o(\omega_s)c}E_{\omega_s}[
\chi^{(3)}_{\rm SPM}|E_{\omega_s}|^2+2\chi^{(3)}_{\rm XPM}|E_{\omega_p}|^2
]
=0
\end{multline}
where $n_o$ is the $o$ polarized refractive index, modelled by the BBO Sellmeier equations reported in \cite{Zhang:2000}. Here we have only included the type I $oo\rightarrow e$ SHG, SFG and DFG possibilities, all having the same effective nonlinearity
\begin{align}\label{eq:deff}
d_{\rm eff}(\theta,\phi)=d_{31}\sin\theta-d_{22}\cos\theta\sin 3\phi
\end{align}
For the cut we use this is optimized for $\phi=-\pi/2$ as $d_{22}/d_{31}<0$ \cite{Bache:2013}, and this implies that $d_{\rm eff}=0$ for the type II interaction $oe\rightarrow e$ and the type 0 interaction $oo\rightarrow o$. For the cubic terms only included the XPM terms that involve the strong soliton and probe fields, and not the weak $e$-polarized SHG, SFG and DFG modes. We also note that even if BBO is anisotropic then the XPM term between two $o$-polarized modes is the same as the SPM term, i.e. $\chi^{(3)}_{\rm XPM}=\chi^{(3)}_{\rm SPM}=c_{11}$ \cite{Bache:2013}. In this identity Miller's scaling is neglected, as it actually is in our code (a single frequency-independent $\chi_{jk}^{(3)}$ is used for each tensor element \cite{Guo:2013}). However, for clarity let us keep them separated, as one could imagine cases where the probe is not the same polarization as the soliton. For the probe we equivalently have
\begin{multline}\label{eq:probe}
id_z E_{\omega_p}+\frac{\omega_p d_{\rm eff}}{cn_o(\omega_p)}[
E_{\omega_p}^*E_{2\omega_p}e^{i\Delta k_{\omega_p}^{\rm SHG}z}\\
+
E_{\omega_s}E_{\omega_s-\omega_p}^*e^{i\Delta k_{\omega_s-\omega_p}^{\rm DFG}z}
\\
+E_{\omega_s}^*E_{\omega_s+\omega_p}e^{i\Delta k_{\omega_s+\omega_p}^{\rm SFG}z}
]
\\+
\frac{3\omega_s}{8n_o(\omega_p)c}E_{\omega_p}[
\chi^{(3)}_{\rm SPM}|E_{\omega_p}|^2+2\chi^{(3)}_{\rm XPM}|E_{\omega_s}|^2
]
=0
\end{multline}
The phase-mismatch coefficients are
\begin{align}
\label{eq:Dks-SHG}
\Delta k_{\omega_j}^{\rm SHG}(\theta)&=k_e(2\omega_j,\theta)-2k_o(\omega_j)\\
\label{eq:Dks-SFG}
\Delta k_{\omega_s+\omega_p}^{\rm SFG}(\theta)&=k_e(\omega_s+\omega_p,\theta)-k_o(\omega_s)-k_o(\omega_p)\\
\label{eq:Dks-DFG}
\Delta k_{\omega_s-\omega_p}^{\rm DFG}(\theta)&=k_o(\omega_s)-k_o(\omega_p)-k_e(\omega_s-\omega_p,\theta)
\end{align}
where $k_e(\omega,\theta)=n_e(\omega,\theta)\omega/c$ and as per usual $n_e(\omega,\theta)= [\cos^2\theta/n_o^2(\omega)+\sin^2\theta/n_e^2(\omega)]^{-1/2}$ where $n_e(\omega)$ is the BBO $e$-polarized refractive index. In what follows we drop the explicit dependence of the $\Delta k$'s and $d_{\rm eff}$ on $\theta$ and $\phi$.

Let us now write the basic plane-wave equations for the SHG, SFG and DFG $e$-polarized modes, disregarding irrelevant quadratic contributions (i.e. the $ee\rightarrow e$ type 0 interaction, which is very weak since it will be heavily phase mismatched and the relevant $d_{\rm eff}$ \cite[Eq. (6)]{Bache:2013} is not as high as $d_{33}$ in, e.g., LiNbO$_3$), cubic nonlinearities (which we assume to a very good approximation to be irrelevant for the harmonics, as the phase mismatch is larger and thus their intensities too low) and chromatic dispersion (we are only interested in the nonlinear terms at the moment).
\begin{align}
\label{eq:SHG}
id_z E_{2\omega_j}+\frac{2\omega_j d_{\rm eff}}{cn_e(\omega_j,\theta)}
\tfrac{1}{2}E_{\omega_j}^2e^{-i\Delta k_{\omega_j}^{\rm SHG}z}=0
\\
\label{eq:DFG}
id_z E_{\omega_s-\omega_p}+\frac{(\omega_s-\omega_p) d_{\rm eff}}{cn_e(\omega_s-\omega_p,\theta)}
E_{\omega_s}E_{\omega_p}^*e^{i\Delta k_{\omega_s-\omega_p}^{\rm DFG}z}=0
\\
\label{eq:SFG}
id_z E_{\omega_s+\omega_p}+\frac{(\omega_s+\omega_p) d_{\rm eff}}{cn_e(\omega_s+\omega_p,\theta)}
E_{\omega_s}E_{\omega_p}e^{-i\Delta k_{\omega_s+\omega_p}^{\rm SFG}z}=0
\end{align}
With the usual cascading ansatz $\Delta k L\gg 2\pi$ we can find the harmonic fields in the cascading limit \cite{bache:2007a}
\begin{align}
\label{eq:SHG-casc}
E_{2\omega_j}=-\frac{\omega_j d_{\rm eff}}{cn_e(\omega_j,\theta)\Delta k_{\omega_j}^{\rm SHG}}
E_{\omega_j}^2e^{-i\Delta k_{\omega_j}^{\rm SHG}z}
\\
\label{eq:DFG-casc}
E_{\omega_s-\omega_p}=\frac{(\omega_s-\omega_p) d_{\rm eff}}{cn_e(\omega_s-\omega_p,\theta)\Delta k_{\omega_s-\omega_p}^{\rm DFG}}
E_{\omega_s}E_{\omega_p}^*e^{i\Delta k_{\omega_s-\omega_p}^{\rm DFG}z}
\\
\label{eq:SFG-casc}
E_{\omega_s+\omega_p}=-\frac{(\omega_s+\omega_p) d_{\rm eff}}{cn_e(\omega_s+\omega_p,\theta)\Delta k_{\omega_s+\omega_p}^{\rm SFG}}
E_{\omega_s}E_{\omega_p}e^{-i\Delta k_{\omega_s+\omega_p}^{\rm SFG}z}
\end{align}
When plugging these into Eqs. (\ref{eq:soliton})-(\ref{eq:probe}) we get
\begin{align}\label{eq:soliton-NLSE}
id_z E_{\omega_s}&+
\frac{3\omega_s}{8n_o(\omega_s)c}E_{\omega_s}[
\chi^{(3)}_{\rm eff, SPM}(\omega_s)|E_{\omega_s}|^2\\
\nonumber
&+2\chi^{(3)}_{\rm eff, XPM}|E_{\omega_p}|^2
]
=0
\\\label{eq:probe-NLSE}
id_z E_{\omega_p}&+
\frac{3\omega_p}{8n_o(\omega_p)c}E_{\omega_p}[
\chi^{(3)}_{\rm eff, SPM}(\omega_p)|E_{\omega_p}|^2\\
\nonumber
&+2\chi^{(3)}_{\rm eff, XPM}|E_{\omega_s}|^2
]
=0
\end{align}
Here
\begin{align}
\label{eq:chi3-eff-SPM}
\chi^{(3)}_{\rm eff, SPM}(\omega_j)=&\chi^{(3)}_{\rm SPM}+
\chi^{(3),{\rm SHG}}_{\rm casc}(\omega_j)
\\
\label{eq:chi3-eff-XPM}
\chi^{(3)}_{\rm eff, XPM}=&\chi^{(3)}_{\rm XPM}+
\chi^{(3),\rm SFG}_{\rm casc}(\omega_s+\omega_p)
\\\nonumber
&+\chi^{(3),{\rm DFG}}_{\rm casc}(\omega_s-\omega_p)
\end{align}
where
\begin{align}
\label{eq:chi3-SHG}
\chi^{(3),\rm SHG}_{\rm casc}(\omega_j)=
-\frac{8\omega_jd_{\rm eff}^2}{3cn_e(2\omega_j,\theta)\Delta k_{\omega_j}^{\rm SHG}}
\\
\label{eq:chi3-SFG}
\chi^{(3),\rm SFG}_{\rm casc}(\omega_s+\omega_p)=
-\frac{4(\omega_s+\omega_p)d_{\rm eff}^2}
{3cn_e(\omega_s+\omega_p,\theta)\Delta k_{\omega_s+\omega_p}^{\rm SFG}}
\\
\label{eq:chi3-DFG}
\chi^{(3),\rm DFG}_{\rm casc}(\omega_s-\omega_p)=
\frac{4(\omega_s-\omega_p)d_{\rm eff}^2}
{3cn_e(\omega_s-\omega_p,\theta)\Delta k_{\omega_s-\omega_p}^{\rm DFG}}
\end{align}
In the field normalized to the intensity case we have $A_{\omega_j}=E_{\omega_j}\sqrt{2/\varepsilon_0 cn_o(\omega_j)}$, $j=s,p$ and
\begin{align}\label{eq:soliton-NLSE-I}
id_z A_{\omega_s}&+
\frac{\omega_s}{c}A_{\omega_s}[
n_{2,\rm eff}^{\rm SPM}(\omega_s)|A_{\omega_s}|^2
\\\nonumber&
+2n_{2,\rm eff}^{\rm XPM}|A_{\omega_p}|^2
]
=0
\\\label{eq:probe-NLSE-I}
id_z A_{\omega_p}&+
\frac{\omega_p}{c}A_{\omega_p}[
n_{2,\rm eff}^{\rm SPM}(\omega_p)|A_{\omega_p}|^2\\
\nonumber
&+2n_{2,\rm eff}^{\rm XPM}|A_{\omega_s}|^2
]
=0
\end{align}
where
\begin{align}
\label{eq:n2effSPM}
n_{2,\rm eff}^{\rm SPM}(\omega_j)=&
n_{2, \rm Kerr}^{\rm SPM}(\omega_j) +n_{2, \rm casc}^{\rm SHG}(\omega_j)
\\
\label{eq:n2effXPM}
n_{2,\rm eff}^{\rm XPM}=&n_{2, \rm Kerr}^{\rm XPM}
+n_{2, \rm casc}^{\rm SFG}(\omega_s+\omega_p)
+n_{2, \rm casc}^{\rm DFG}(\omega_s-\omega_p)
\end{align}
and
\begin{align}\label{eq:n2SPM}
n_{2, \rm Kerr}^{\rm SPM}(\omega_j)=&\frac{3\chi^{(3)}_{\rm SPM}}{4 \varepsilon_0 c n_o^2(\omega_j)}
\\
\label{eq:n2XPM}
n_{2, \rm Kerr}^{\rm XPM}=&\frac{3\chi^{(3)}_{\rm XPM}}{4 \varepsilon_0 c n_o(\omega_s)n_o(\omega_p)}
\end{align}
while the cascading contributions are
\begin{align}
\label{eq:n2casc-SHG}
&n_{2,\rm casc}^{\rm SHG}(\omega_j)=
-\frac{2\omega_jd_{\rm eff}^2}{\varepsilon_0 c^2 n_o^2(\omega_j)n_e(2\omega_j,\theta)\Delta k_{\omega_j}^{\rm SHG}}
\\
\label{eq:n2casc-SFG}
&n_{2,\rm casc}^{\rm SFG}(\omega_s+\omega_p)=
\\\nonumber
&-\frac{(\omega_s+\omega_p)d_{\rm eff}^2}{\varepsilon_0 c^2 n_o(\omega_s)n_o(\omega_p)n_e(\omega_s+\omega_p,\theta)\Delta k_{\omega_s+\omega_p}^{\rm SFG}}
\\
\label{eq:n2casc-DFG}
&n_{2,\rm casc}^{\rm DFG}(\omega_s-\omega_p)=
\\\nonumber
&\frac{(\omega_s-\omega_p)d_{\rm eff}^2}{\varepsilon_0 c^2 n_o(\omega_s)n_o(\omega_p)n_e(\omega_s-\omega_p,\theta)\Delta k_{\omega_s-\omega_p}^{\rm DFG}}
\end{align}


\begin{figure}[tb]
\includegraphics[width =\linewidth]{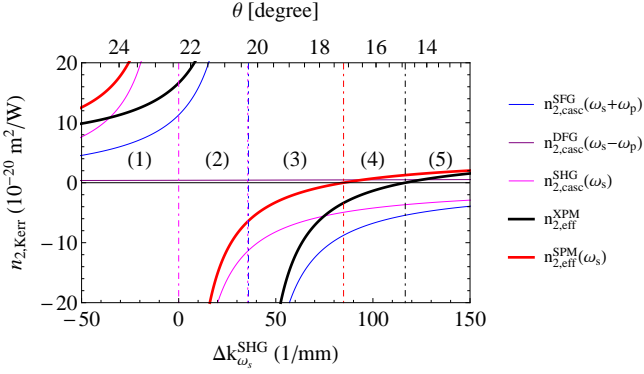}
\caption{The induced cascaded nonlinear refractive indices vs. the SHG phase-mismatch parameter (controlled by the crystal angle $\theta$), calculated for $\lambda_s=1.1\mic$ and $\lambda_p=1.65\mic$. The domains are divided as follows: (1+4+5) Effective self-focusing SPM regimes ($n_{2, \rm eff}^{\rm SPM}(\omega_s)>0$), so solitons cannot be excited; (2+3) Effective self-defocusing SPM regimes ($n_{2, \rm eff}^{\rm SPM}(\omega_s)<0$), i.e. soliton regimes; (1+2+5) positive XPM regimes ($n_{2, \rm eff}^{\rm SPM}(\omega_s)>0$); (3+4) negative XPM regimes ($n_{2, \rm eff}^{\rm SPM}(\omega_s)<0$); (1+2) resonant nonlocal cascaded SHG regimes; (3+4+5) non-resonant nonlocal cascaded SHG regimes.
}
\label{fig:Dk-n2}
\end{figure}

A requirement for the soliton to "reflect" the probe is that the potential imposed by the soliton on the probe is scattering, i.e. a barrier. As the probe is launched in the anomalous dispersion regime, a scattering potential requires that the XPM effective nonlinear index of the probe ($n_{2, \rm eff}^{\rm XPM}$) is negative, i.e. self-defocusing. As we showed above this has three contributions: $n_{2, \rm eff}^{\rm XPM}=n_{2, \rm Kerr}^{\rm XPM}+n_{2, \rm casc}^{\rm SFG}(\omega_s+\omega_p)+n_{2, \rm casc}^{\rm DFG}(\omega_s-\omega_p)$, viz. the material Kerr XPM as well as cascaded SFG and DFG between the soliton and probe.

In Fig. \ref{fig:Dk-n2} we show how the crystal tuning angle $\theta$ controls the various regimes; the self-defocusing soliton can only be excited in regimes (2+3). In turn the effective XPM term is only negative in regimes (3+4); at the boundary to regime (2) it becomes phase matched, and thus in (1+2) it is positive. Coincidentally the boundary between regimes (2) and (3) also marks the transition where the cascaded SHG becomes non-resonant, which it is in regime (3+4+5). In the non-resonant regime the cascading is ultrabroadband and induces minimal self-steepening on the soliton \cite{bache:2007a,zhou:2012,zhou:PhysRevA.90.013823}.

Thus, for the wavelengths and tuning angles considered here, $n_{2, \rm casc}^{\rm DFG}(\omega_s-\omega_p)$ is negligible due to a large phase mismatch. Therefore the XPM scattering potential sign and magnitude is by and large a competition between the material Kerr XPM effect and the cascaded SFG effect. 



%

\end{document}